# Trapping of dark vector solitons in a fiber laser

**H. Zhang[1], D. Y. Tang[1]*, L. M. Zhao[1], X. Wu[1]**

**Q. L Bao[2] and K. P. Loh[2]**

[1]School of Electrical and Electronic Engineering,

Nanyang Technological University, Singapore 639798

[2]Department of Chemistry, National University of Singapore, 3 Science Drive 3,

Singapore 117543

*: edytang@ntu.edu.sg, corresponding author.

We report on the experimental observation of incoherently coupled dark vector soliton trapping in a fiber laser. Dark vector solitons along the two orthogonal polarization directions though with large difference in central frequency, energy and darkness could be still synchronously trapped as a group velocity locked dark vector soliton in virtue of the incoherent interactions. Our numerical simulation could well rebirth the experimental observations and confirm the existence of incoherently coupled dark vector soliton under certain conditions.

Optical soliton, in the form of a localized coherent structure, has been the object of intensive theoretical and experimental studies over the last decades [1-3]. It is now well-known that the nonlinear Schrödinger equation (NLSE) could govern the formation of either bright or dark optical solitons depending on the sign of the fiber group velocity dispersion (GVD) [4, 5]. For future photonic computing, optical logic gates and light control light devices ,derived from soliton interaction by virtue of cross-phase modulation (XPM), are attractive, since the general soliton features could be still sustained thanks to the conjunct interactions. It is physically and technically meaningful to explore the essence and attribute of bright or dark soliton interactions. Mutual bright soliton interaction as a result of XPM in optical fiber, also defined as soliton trapping, was first theoretically predicted by Curtis R. Menyuk [6, 7] and then experimentally validated [8-10]. Soliton trapping is denoted as two orthogonally polarized time-delayed optical solitons in birefringence fibers trapping each other and co-propagating jointly as long as their group velocity difference could be compensated through shifting their central frequencies in opposite directions by means of XPM. Kivshar theoretically foretold a novel incoherently coupled dark vector soliton comprised by two grey solitons, belonging to two orthogonal polarization modes, strongly coupled with the help of XPM, in a highly birefringent fiber [11]. However, to the best of our knowledge, no incoherently coupled temporal dark vector soliton formation experiments have been reported.

Recently, we have experimentally achieved the formation of stable dark scalar solitons in an all-normal dispersion fiber laser [12]. It is to expect that through appropriately devising the cavity, dark vector soliton operation could be established if the polarization dependent component is replaced with a polarization insensitive device which profit to stabilize dark solitons. Triggered by facts that dark pulses could steadily exist in a cavity soliton laser in the presence of saturable absorber [13], a weak saturable absorber based on G, was specially introduced into an all normal dispersion fiber laser cavity. In this paper, indeed, stable dark vector soliton could be ultimately observed. Depending on the pumping strength and cavity birefringence, both single and multiple dark vector solitons are realized. Further investigation demonstrates that the dark vector soliton is composed of two orthogonally polarized grey solitons incoherently coupled as a consequence of XPM, which is then numerically affirmed by our simulation model based on NLSEs. We deem that Kivshar's theory on dark vector soliton could be both experimentally and numerically justified.

Our fiber laser is schematically shown in Fig. 1. The laser cavity is an all-normal dispersion fiber ring consisting of 5.0 m, 2880 ppm Erbium-doped fiber (EDF) with group velocity dispersion (GVD) of -32 (ps/nm)/km, and **157.6** m dispersion compensation fiber (DCF) with GVD of -4 (ps/nm)/km. A polarization insensitive isolator together with an in-line polarization controller (PC) was employed in the cavity to force the unidirectional operation of the ring and control the light polarization. A 50% fiber coupler was adopted to output the signal, and the laser was pumped by a high power Fiber Raman Laser source (KPS-BT2-RFL-1480-60-FA) of wavelength 1480 nm. The maximum pump power can reach as high as 5W. All the passive components used (WDM, Coupler, Isolator) were made of the DCF. An optical spectrum analyzer (Ando AQ-6315B) and a 350MHZ oscilloscope (Agilen 54641A) together with two 2GHZ photo-detectors were utilized to simultaneously monitor the spectra and pulse train, respectively.

In order to distinguish the laser emission along the two orthogonal principal polarization directions of the cavity, an in line polarization beam splitter (PBS) combined with an external cavity PCs which was operated to optimize the output polarization state, was spliced after the output coupler so as to split the output signals into two orthogonal polarizations which were then simultaneously monitored by a multi-channeled oscilloscope associated with two identical 2GHZ photo-detectors. In particular, optical and electric path discrepancy between the two polarization-resolved signals should be eliminated as weak as possible. Recently, we have readily obtained scalar dark soliton in a fiber laser with a polarization dependent isolator under appropriate pump strength and negative cavity feedback [12]. But dark vector solitons were brittle and sensitive to environment perturbations if a polarization dependent isolator was used to replace the polarization dependent one. Opportunely, dark vector solitons could be stabilized by means of incorporating a weak saturable absorber, which represents a nonlinear loss term to suppress the dark soliton instabilities, inside the cavity. This weak saturable absorber is specially made of atomic layer G film attached onto fiber pigtails [14]. About 2~3 layers of G covered on the fibre core area is identified through a combination of Raman and contrast spectroscopy. It has a saturable absorption modulation depth of 64.4%, a saturation fluence of 7.1 $\mu J/cm^2$ and a carrier relaxation time 1.67 ps (determined by a time-resolved pump-probe method). Due to the knowledge that G could be readily saturated, its mode locking ability is severely restrained, meaning that the formation of bright soliton turns to be formidable and thus chances of dark soliton formation become higher.

As soliton performances are strongly cavity birefringence and pump strength dependent, the output could be significantly modulated through rotating PCs and varying the pump. In the experiments, laser prefers to operate in the continuous wave (CW) regime, represented by a straight line above ground level in the oscilloscope trace after optical-electrical conversion using a fast photodiode. Moreover, we noticed that when CW beam intensity kept sufficiently strong under a high pump power ~2w, CW instability began to emerge and heavy intensity dips became visible and even could survive over a long period if PCs were precisely adjusted. Carefully examining the polarization resolved oscilloscope traces; it is to note that while along one principal polarization direction the laser emitted a train of intensity dips conjointly with a strong CW background (upper trace of Fig. 2a), along the orthogonal polarization direction reciprocal dark solitonic features was observed (lower trace of Fig. 2a). Anyhow, tuning the external cavity PCs, which physically corresponds to affect the output fiber birefringence and eventually influence the dark soliton extinction ratio after passing through PBS, the two polarization resolved oscilloscope traces always maintained a single dark soliton pulse train despite the relative strength could be extensively adjusted. Further polarization resolved spectra as shown in Fig. 2b illustrates that the two orthogonal polarization components have clearly different spectral distribution and ~0.1nm central wavelength difference, which is comparably large enough in contrast with the 3dB spectral bandwidth for each polarization ~0.5nm, if the central spectra, as shown in the insert of Fig. 2b, is deliberately inspected. Based on the above experimental observation, we could conclude that the observed dark soliton is not scalar (linearly polarized) but vectorial and each polarizations are incoherently coupled to form a group velocity locked dark vector soliton [11]. Akin to the nature of dark scalar soliton [12], multiple dark vector solitons could be viably formed as shown in Fig. 2c, if the pump strength is further increased, and each of the dark solitons does not necessarily have the equivalent shallowness, i.e. the failure of "soliton energy quantization effect". In addition, juxtaposition of the zoom-in oscilloscope traces of dark vector solitons at different positions as shown in Fig. 2d, obviously, it turns up that the relative shallowness ratio of each polarization could be unrelated, revealing that the darkness for each polarization is neither invariant nor "quantized" but fluctuant along the propagation. It could be traced back to the formation mechanism of dark soliton. Kivshar prognosticated that any dip on the CW background pulse of one polarization would instantly breed a similar dip in its orthogonal polarization mode incoherently coupled to the maternal pulse. Accordingly, as individual dark soliton arises from the weak perturbation of CW and thus pertains to their initial "seed dips", the uncorrelated formation renders the arbitrary darkness for each polarization. Fig. 2d also shows that the two polarizations are temporally synchronized and have almost the same pulse profiles and widths that are limited by the resolution of the electronic detection system. Due to the low pulse repetition rate, the pulse width of the dark solitons could not be measured with the autocorrelator.

To confirm whether incoherently coupled dark solitons could be formed under the current cavity parameters; we numerically simulated the operation of the laser with the standard split-step Fourier technique to solve the equations and a so-called pulse tracing method to model the effects of laser oscillation [15]. We used the following coupled Ginzburg-Landau equations to describe the pulse propagation in the weakly birefringent fibers:

$$\begin{cases} \frac{\partial u}{\partial z} = i\beta u - \delta\frac{\partial u}{\partial t} - \frac{ik''}{2}\frac{\partial^2 u}{\partial t^2} + \frac{ik'''}{6}\frac{\partial^3 u}{\partial t^3} + i\gamma(|u|^2 + \frac{2}{3}|v|^2)u + \frac{i\gamma}{3}v^2u^* + \frac{g}{2}u + \frac{g}{2\Omega_g^2}\frac{\partial^2 u}{\partial t^2} \\ \frac{\partial v}{\partial z} = -i\beta v + \delta\frac{\partial v}{\partial t} - \frac{ik''}{2}\frac{\partial^2 v}{\partial t^2} + \frac{ik'''}{6}\frac{\partial^3 v}{\partial t^3} + i\gamma(|v|^2 + \frac{2}{3}|u|^2)v + \frac{i\gamma}{3}u^2v^* + \frac{g}{2}v + \frac{g}{2\Omega_g^2}\frac{\partial^2 v}{\partial t^2} \end{cases} \quad (1)$$

Where, u and v are the normalized envelopes of the optical pulses along the two orthogonal polarized modes of the optical fiber. $2\beta = 2\pi\Delta n/\lambda = 2\pi/L_b$ is the wave-number difference between the two modes. $2\delta = 2\beta\lambda/2\pi c$ is the inverse group velocity difference. $k''$ is the second order dispersion coefficient, $k'''$ is the third order dispersion coefficient and $\gamma$ represents the nonlinearity of the fiber. $g$ is the saturable gain coefficient of the fiber and $\Omega_g$ is the bandwidth of the laser gain. For undoped fibers $g$=0; for erbium doped fiber, we considered its gain saturation as

$$g = G\exp[-\frac{\int(|u|^2 + |v|^2)dt}{P_{sat}}] \quad (2)$$

where $G$ is the small signal gain coefficient and $P_{sat}$ is the normalized saturation energy. Similar to SESAM, the saturable absorption of G could be also described by the rate equation [16]:

$$\frac{\partial l_s}{\partial t} = -\frac{l_s - l_0}{T_{rec}} - \frac{|u|^2 + |v|^2}{E_{sat}} l_s \quad (3)$$

Where $T_{rec}$ is the absorption recovery time, $l_0$ is the initial absorption of the absorber, and $E_{sat}$ is the absorber saturation energy. To make the simulation possibly close to the experimental situation, we used the following parameters: $\gamma$=3 $W^{-1}km^{-1}$, $\Omega_g$ =16 nm, $P_{sat}$=50 pJ, $k''_{DCF}$= 4.6 $ps^2$/km, $k''_{EDF}$= 41.6 $ps^2$/km, $k'''$= -0.13 $ps^3$/km, $E_{sat}$=1 pJ, $l_0$=0.5, and $T_{rec}$ = 2 ps, Cavity length L= 162.6 m.

To faithfully simulate the creation of incoherently coupled dark soliton, we always use an arbitrary weak pulse as the input initial condition and the weak saturable absorber functions in the transmission mode rather than reflection mode. Stable dark vector soliton could be conceivably obtained under diverse cavity birefringence regimes. Fig. 3 and Fig. 4 shows the evolution of dark vector soliton within both time and frequency domain when the cavity beat length was chosen as Lb=L/40 and Lb=L/60, respectively. Numerically, we found that an antiphase type of periodic intensity variation between CW background inherent to the two orthogonally polarized dark solitons took place under weak cavity birefringence, for example, Lb=100L. However, the stronger the linear cavity birefringence the weaker such intensity variation, eventually, coherent energy exchange vanished as seen in Fig. 3a. Due to strong cavity birefringence where phase matching conditions for four wave mixing are challenging to fulfill [17], coherent coupling between the two polarization components of a dark vector soliton become vain while incoherent interaction arising from XPM still militates. This can explain why the two dark polarization components could be still synchronously entrapped in temporal domain and have little central frequency difference, as shown in Fig.3, even though the cavity birefringence Lb=L/40 is very strong. Further strengthening the cavity birefringence, the two polarization components could be still concurrently bounded and propagates as one non-dispersive unit but they are temporally delayed, as shown in Fig. 4b when Lb=L/60. Moreover, their central frequencies have 0.1nm difference, matching well with our experimental observation. Whereas, under further higher cavity birefringence, the two dark soliton could no longer be trapped as one entity but break up and spread at their respective group velocities, indicating that a larger time delay correspond to a less effective attraction, and consequently uncoupled [11]. Finally, we note that in the current simulation model, the dips of the numerically calculated dark solitons could reach zero under weak cavity birefringence but are nonzero as shown in Fig. 3b and Fig. 4b, implying that they are grey solitons. CW backgrounds for each grey soliton are different, which confirms Kivshar's prediction that two grey solitons with different background intensities could be incoherently coupled by virtue of XPM. Furthermore, grey solitons in Fig. 4b are slightly asymmetric near the central dips but converge to consistent asymptotic values provided that $t \to \pm\infty$. We conjecture that such asymmetry is a natural consequence of laser cavity effect, which does not exist in Kivshar's model, because apart from the balance of normal cavity dispersion and fiber nonlinear Kerr effect, a dark vector soliton is also subject to the cavity gain and losses, boundary condition, as well as the filtering effect caused by the strong cavity birefringence. Nevertheless, our numerical simulation model could well explain the formation of incoherently coupled dark vector soliton trapping in a fiber laser.

In conclusion, incoherently coupled dark vector soliton has been experimentally and numerically observed in an all normal dispersion fiber laser where a weak saturable absorber based on Graphene is purposely introduced to cooperatively stabilize the dark vector soliton. We believe our study could not only perfectly confirm Kivshar's theory but also verify the feasibility of dark soliton trapping technology. Although the dark soliton trapping is achieved in a fiber laser, due to the general applicability of NLSE, dark soliton trapping could be a universal hallmark of all coupled dark solitary waves, such as trapping of dark matter wave soliton in Bose-Einstein condensate.

## Figure captions:

Fig. 1: Schematic of the experimental setup. EDF: Erbium doped fiber. WDM: wavelength division multiplexer. DCF: dispersion compensation fiber. PC: polarization controller.

Fig. 2: Polarization resolved (a): oscilloscope trace of single dark vector soliton in the cavity and (b) its corresponding optical spectra Insert: zoom in of (b) near the spectral center. (c) Polarization resolved oscilloscope trace of multiple dark vector soliton in the cavity. (d) Enlarge scale of (c) at different positions.

Fig. 3: Stable dark vector soliton state numerically calculated. (a) Evolution of the dark vector soliton with cavity roundtrips. (b) Zoom in of (a). (c) The corresponding spectra and insert: zoom in of (c). Gain=1500. L/Lb=40.

Fig. 4: Stable dark vector soliton state numerically calculated. (a) Evolution of the dark vector soliton with cavity roundtrips. (b) Zoom in of (a). (c) The corresponding spectra and insert: zoom in of (c). Gain=1500. L/Lb=60.

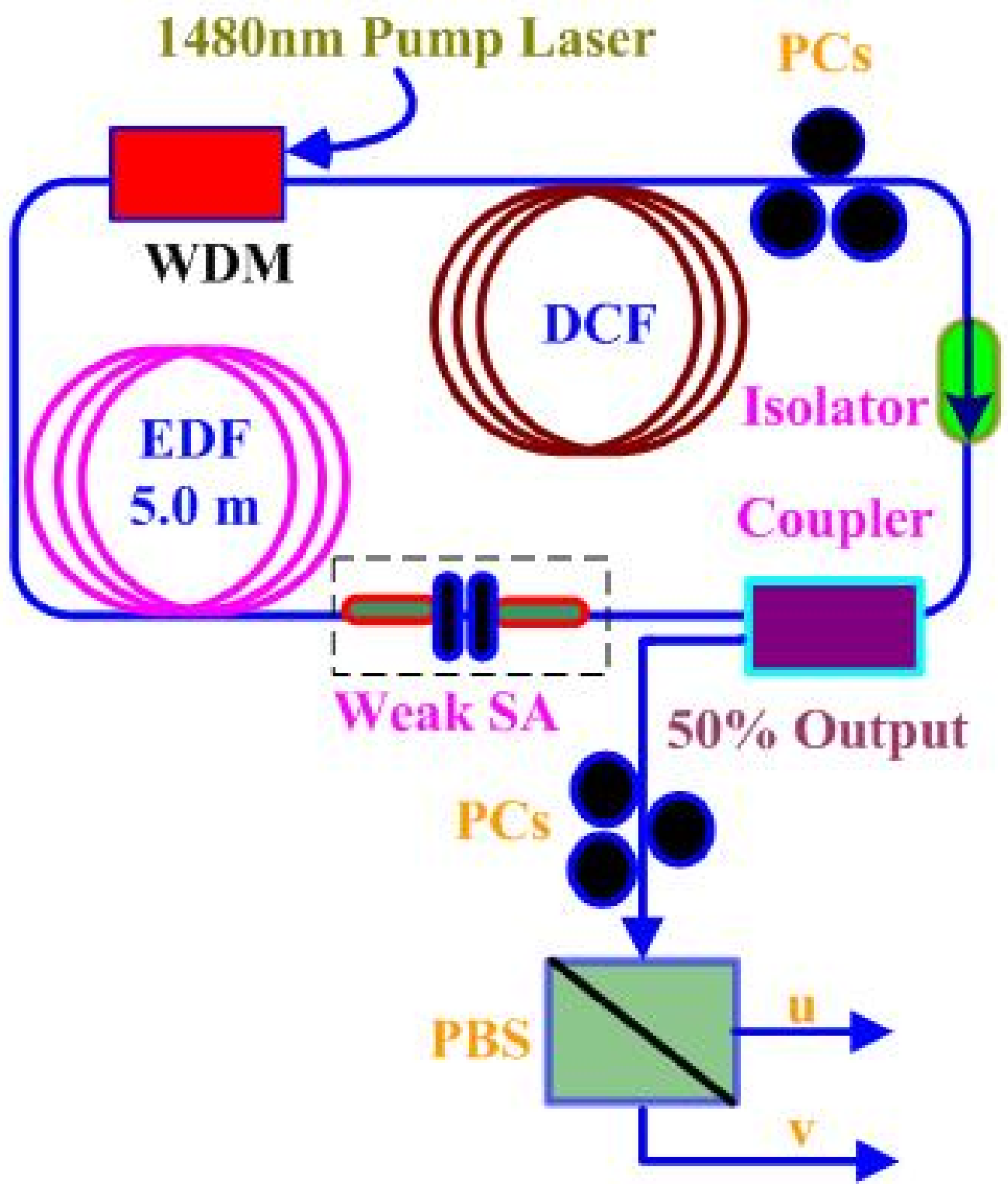

Fig.1 H. Zhang et al

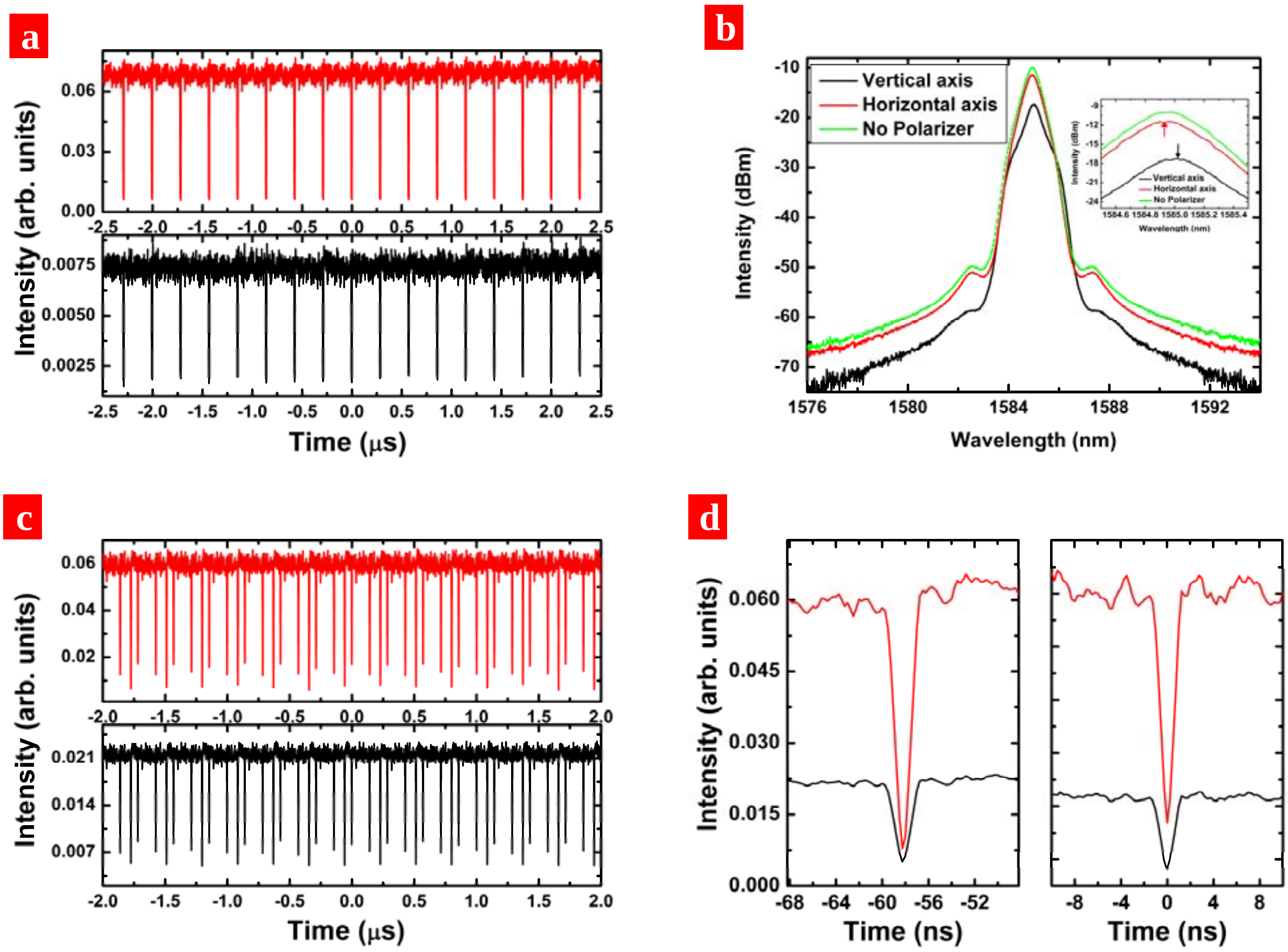

Fig.2 H. Zhang et al

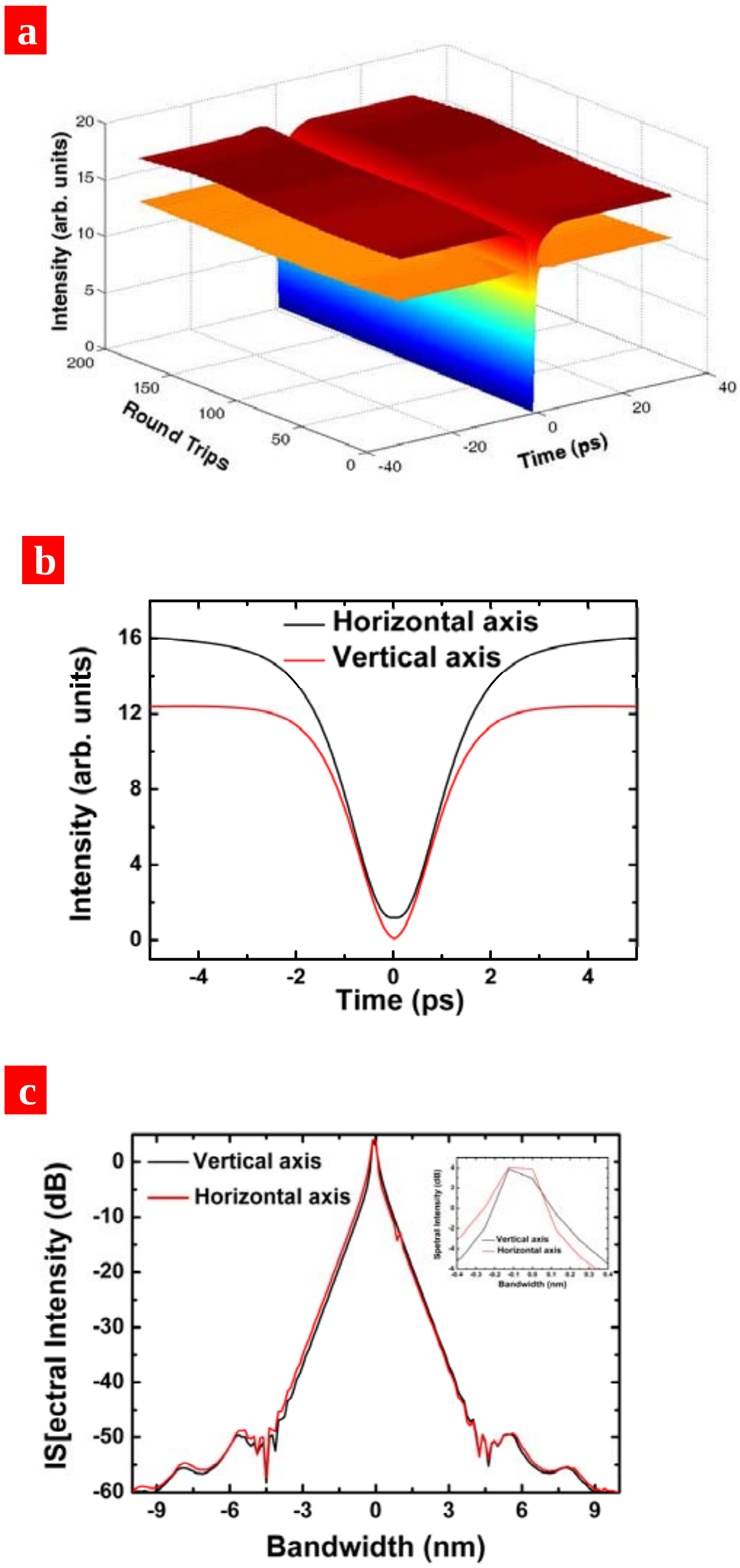

Fig.3 H. Zhang et al

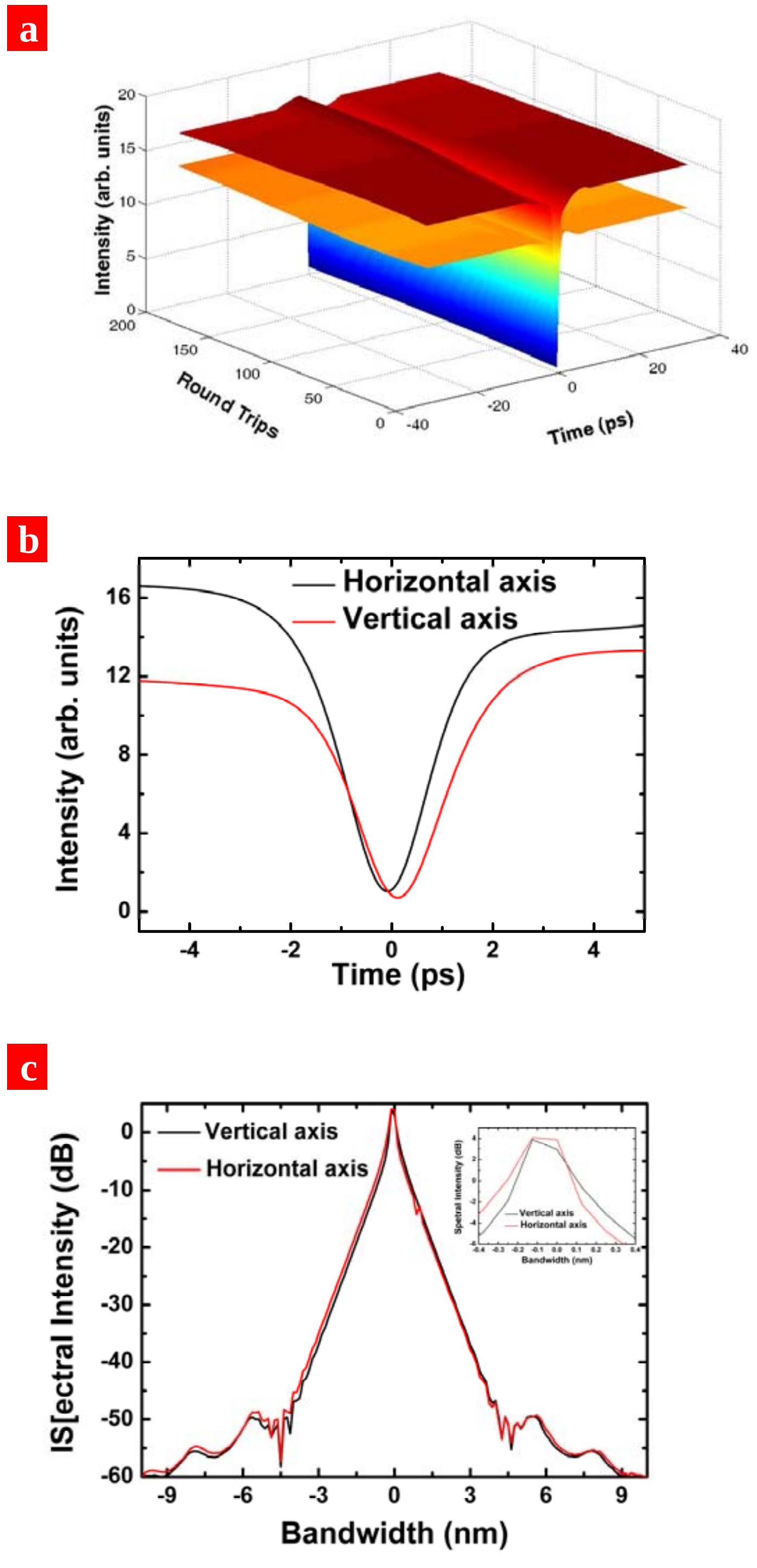

Fig.4 H. Zhang et al